\documentclass[twocolumn,shortnote]{jpsj2}
\def\runauthor{Akira \textsc{Oosawa} {\it et al.}}

\title{Magnetic Excitations in the Quasi-1D Ising-like Antiferromagnet TlCoCl$_3$}

\author{Akira \textsc{Oosawa}\thanks{E-mail address: osawa-a@sophia.ac.jp}, Yoichi \textsc{Nishiwaki}$^{1}$, Tetsuya \textsc{Kato}$^{2}$ and Kazuhisa \textsc{Kakurai}$^{3}$}

\inst{Department of Physics, Sophia University, 7-1 Kioi-cho, Chiyoda-ku, Tokyo 102-8554, Japan\\
$^1$Department of Physics, Tokyo Institute of Technology, Oh-okayama, Meguro-ku, Tokyo 152-8551, Japan\\
$^2$Faculty of Education, Chiba University, 1-33 Yayoi-cho, Inage-ku, Chiba 273-8522, Japan\\
$^3$Advanced Science Research Center, Japan Atomic Energy Research Institute, Tokai, Ibaraki 319-1195, Japan}

\recdate{\today}

\abst{Neutron inelastic scattering measurements have been performed in order to investigate the magnetic excitations in the quasi-1D Ising-like antiferromagnet TlCoCl$_3$. We observed the magnetic excitation, which corresponds to the spin-wave excitation continuum corresponding to the domain-wall pair excitation in the 1D Ising-like antiferromagnet. According to the Ishimura-Shiba theory, we analyzed the observed spin-wave excitation, and the exchange constant $2J$ and the anistropy $\epsilon$ were estimated as 14.7~meV and 0.14 in TlCoCl$_3$, respectively.}  

\kword{TlCoCl$_3$, quasi-1D Ising-like antiferromagnet, magnetic excitation, excitation continuum, domain-wall, neutron inelastic scattering}

\begin{document}
\sloppy
\maketitle

TlCoCl$_3$ has the CsNiCl$_3$-type hexagonal crystal structure (space group symmetry $P6_3/mmc$) at room temperature, in which the magnetic Co ions form the linear chain along the $c$-axis and these chains make triangular lattice in the $c$-plane. Because Co ions have a large Ising anistropy along the $c$-axis due to the Kramers doublet, this system can be expressed as the quasi-1D Ising-like antiferromagnet. From the dielectric constant measurements, it was found that this system undergoes the successive structural phase transitions with ferroelectricity at $T_{\rm st2}=165$~K, $T_{\rm st3}=75$~K and $T_{\rm st4}=68$~K \cite{Nishiwaki1}. Besides, it was found from the magnetization measurements that the magnetic phase transition occurs at $T_{\rm N}$=29.5~K \cite{Nishiwaki1}. \par
Recently, the neutron diffraction measurements have been carried out in this system in order to investigate the crystal and magnetic structures. It was found that the crystal structure varies from the hexagonal structure ($P6_3/mmc$) at room tempearture to the orthorombic one ($Pbca$) below $T_{\rm st4}$ and the magnetic Bragg reflections indicative of the {\it up-up-down-down}-type magnetic structure have been observed below $T_{\rm N}$ at ${\pmb Q}=(\frac{h}{8}, \frac{h}{8}, l)$ with odd $h$ and odd $l$, which are assigned by the indices of the room-temperature hexagonal lattice \cite{Nishiwaki2}. In this short note, we report the results of the first neutron inelastic scattering experiments in order to investigate the magnetic excitations in this system. \par
Neutron inelastic scattering experiments were performed on TlCoCl$_3$ using the JAERI-TAS2 spectrometer installed at JRR-3M, Tokai, Japan. The constant-$k_{\rm f}$ mode was taken with a fixed final neutron energy $E_{\rm f}$ of 14.7~meV. Due to the experimental limitation, the maximum of the energy transfer $\Delta E$ is 18~meV. A pyrolitic graphite-filter was used to suppress the higher order contaminations. The collimations were set as 14'-80'-80'-80'. We used a single crystal of TlCoCl$_3$ with a volume of approximately 0.5~cm$^3$. The samples were mounted in a cryostat with its (1, 1, 0) and (0, 0, 1) axes in the scattering plane. The indices of the room-temperature hexagonal lattice are used in the present experiments. The relations between the indices of the room-temperature hexagonal lattice and those of the low-temperature orthogonal lattice are shown in ref. \citen{Nishiwaki2}. The crystallographic parameters were determined as $a = 6.83$~${\rm \AA}$ and $c = 5.97$~${\rm \AA}$ at $T=10$~K, which are defined in the room-temperature hexagonal lattice. \par
\begin{figure}[t]
\begin{center}
\includegraphics[width=60mm]{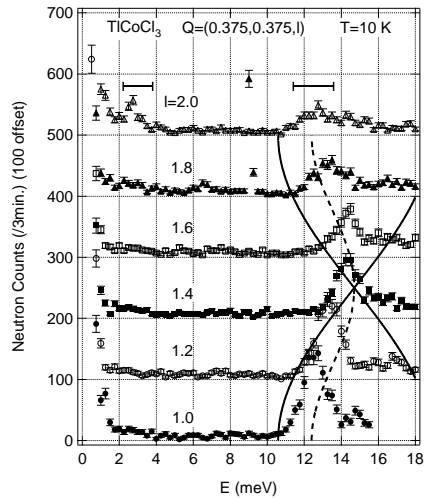}%
\end{center}
\vspace{-0.5cm}
\caption{Profiles of the constant-${\pmb Q}$ energy scans in TlCoCl$_3$ for ${\pmb Q}$ along (0.375, 0.375, $l$) at $T=10$~K. The dotted and solid lines denote the fitting results of the ${\pmb Q}$-dependence of the peak position (eq. (\ref{B}) ) and the boundaries of the continuum (eq. (\ref{C}) ) with $2J=14.7$~meV and $\epsilon=0.14$ derived from the IS theory, respectively. The horizontal error bars indicate the calculated resolution widths at $E=3$ and 12.5 meV. \label{Fig1}}
\end{figure}  
Figure \ref{Fig1} shows the profiles of the constant-${\pmb Q}$ energy scans in TlCoCl$_3$ for ${\pmb Q}$ along (0.375 (=$\frac{3}{8}$), 0.375, $l$) at $T=10$~K. As shown in Fig. \ref{Fig1}, a excitation with a little dispersive and asymmetric was observed at $E \sim 13$~meV. Such excitation has been also observed in the quasi-1D Ising-like antiferromagnet CsCoCl$_3$ \cite{Yoshizawa,Nagler}. Previously, the magnetic excitations have been theoretically studied in the 1D Ising-like antiferromagnet with the following Hamiltonian (\ref{A}) by Ishimura and Shiba (IS) \cite{Ishimura}, 
\begin{equation}
\label{A}
{\cal H} = 2J \sum_j \left[ S_j^z S_{j+1}^z + \epsilon \left( S_j^x S_{j+1}^x + S_j^y S_{j+1}^y\right) \right]
\end{equation}
in order to compare the magnetic excitations in CsCoCl$_3$. Note that the additional terms, such as the exchange mixing \cite{Goff,Nagler} and the next-nearest-neighbor ferromagnetic interaction \cite{Matsubara,Shiba2}, to eq. (\ref{A}) were discussed after the IS theory. However, the IS theory is enough to express the main features of magnetic excitations in CsCoCl$_3$ so that we compare the obtained results of TlCoCl$_3$ with the IS theory below. In the IS theory, the spin-wave excitation continuum corresponding to the domain-wall pair excitation was found. This excitation is a little dispersive and have the asymmetrical spectrum due to the continuum. This feature corresponds to the observed excitation of TlCoCl$_3$, therefore this infer that the observed excitation corresponds to the domain-wall pair excitation along the $c$-axis. The IS theory also predicted the existence of the central mode corresponding to the motion of thermally activated domain walls firstly predicted by Villain \cite{Villain}. As shown in Fig. \ref{Fig1}, a little peak structure can be seen at $E \sim 3$~meV. Because the intensity of the central mode decreases with decreasing temperature and almost vanishes at low temperature \cite{Ishimura}, the observed peak structure may correspond to the central mode. Note that the present measurements were carried out below $T_{\rm N}$. At a glance, this attempt causes some inconsistencies for the 1D magnetic excitations, however, it can be expected from the previous measurements on CsCoCl$_3$ that the three-dimensional magnetic ordering affects only spectral shape of the magnetic excitations, hence we can rather expect that the peak intensity increases because the asymmetrical spectral shape becomes more pronounced so that we performed the present experiments below $T_{\rm N}$ in order to gain intensities. In fact, such increase was observed, as shown in Fig. \ref{Fig2} (see below). The double peak structure of the spin-wave excitation, as shown in Fig. \ref{Fig1}, may be due to the quantization of the excitation continuum below $T_{\rm N}$, as discussed by Shiba \cite{Shiba}. \par
\begin{figure}[t]
\begin{center}
\includegraphics[width=50mm]{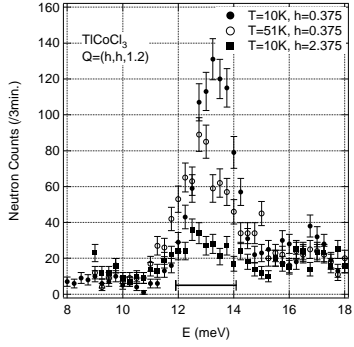}%
\end{center}
\vspace{-0.5cm}
\caption{Profiles of the constant-${\pmb Q}$ energy scans in TlCoCl$_3$ at ${\pmb Q} = (0.375, 0.375, 1.2)$ and (2.375, 2.375, 1.2) for $T=10$ and 51~K. The horizontal error bar indicates the calculated resolution width at $E=13$ meV.\label{Fig2}}
\end{figure}
Figure \ref{Fig2} shows the profiles of the constant-${\pmb Q}$ energy scans in TlCoCl$_3$ at ${\pmb Q} = (0.375, 0.375, 1.2)$ and (2.375, 2.375, 1.2) for $T=10$ and 51~K. As shown in Fig. \ref{Fig2}, the intensity of the spin-wave excitation decreases both with increasing temperature and wave vector ${\pmb Q}$. This indicates that the origin of the observed excitation is of magnetic, and not phonon. \par
\begin{figure}[t]
\begin{center}
\includegraphics[width=50mm]{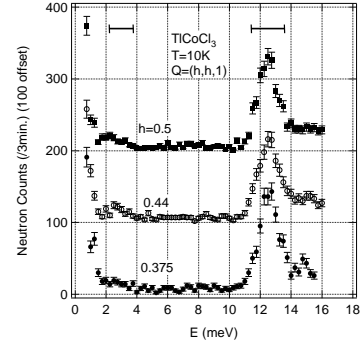}%
\end{center}
\vspace{-0.5cm}
\caption{Profiles of the constant-${\pmb Q}$ energy scans in TlCoCl$_3$ for ${\pmb Q}$ along ($h$, $h$, $1$) at $T=10$~K. The horizontal error bars indicate the calculated resolution widths at $E=3$ and 12.5 meV. \label{Fig3}}
\end{figure}

Figure \ref{Fig3} shows the profiles of the constant-${\pmb Q}$ energy scans in TlCoCl$_3$ for ${\pmb Q}$ along ($h$, $h$, $1$) at $T=10$~K. As shown in Fig. \ref{Fig3}, the spin-wave excitation has no dispersion along the $h$ direction. This means that the spin-wave excitation corresponds to the motion of the domain-wall pair along the chain. This is also consistent with the IS theory. \par
In the IS theory, it was found that the peak position of the spin wave excitation continuum can be expressed as 
\begin{equation}
\label{B}
\omega_q = 2J (1 - 8 \epsilon^2 \cos^2 \pi l)
\end{equation}
and the boundaries of the continuum extending from $\omega_q^-$ to $\omega_q^+$ are given by
\begin{equation}
\label{C}
\omega_q^{\pm} = 2J (1 \pm 2 \epsilon \cos \pi l)
\end{equation}
by means of the perturbation theory from the pure Ising limit. It is noted that these equations may be modified due to the additional terms, as mentioned above. However, the modified analytical equations have not been given so far so that we fit the obtained profiles by these equations. From the fitting, the exchange constant $2J$ and the anistropy $\epsilon$ were obtained as 14.7~meV and 0.14, respectively. The fitting results are shown in Fig. \ref{Fig1}. From these results, we conclude that the observed excitation is the spin-wave excitation continuum corresponding to the domain-wall pair excitation, which is characteristic in the 1D Ising-like antiferromagnet. The obtained exchange constant $2J$ is a little larger than that 12.75~meV of CsCoCl$_3$ \cite{Yoshizawa}, while the anistropy $\epsilon$ is the same as that of CsCoCl$_3$ \cite{Yoshizawa}. This result is consistent with the higher N\'{e}el temperature $T_{\rm N}=29.5$~K of this system than $T_{\rm N}=23$~K of CsCoCl$_3$. \par
For the future problems, it is interesting to investigate the details of the quantization of the excitation continuum below $T_{\rm N}$ by means of the high-resolution neutron scattering experiments \cite{Goff}, the Raman scattering experiments \cite{Shiba,Breitling} and the high-field ESR measurements \cite{Shiba2} because the magnetic structure of TlCoCl$_3$ is different from that of CsCoCl$_3$ \cite{Shiba}. It is also interesting to measure the high-field magnetization process in order to investigate the validity of the eq. (\ref{A}) for TlCoCl$_3$ \cite{Shiba2}. \par
In conclusion, we have presented the results of the neutron inelastic scattering measurements in the quasi-1D Ising-like antiferromagnet TlCoCl$_3$. We observed the magnetic excitation, which corresponds to the spin-wave excitation continuum corresponding to the domain-wall pair excitation in the 1D Ising-like antiferromagnet. According to the Ishimura-Shiba theory, we analyzed the observed spin-wave excitation, and the exchange constant $2J$ and the anistropy $\epsilon$ were estimated as 14.7~meV and 0.14 in TlCoCl$_3$, as shown in Fig. \ref{Fig1}, respectively. \par
This work was supported by the Saneyoshi Scholarship Foundation.


\begin{thebibliography}{99}
\bibitem{Nishiwaki1} Y. Nishiwaki {\it et al.}: J. Phys. Soc. Jpn. {\bf 72} (2003) 2608.
\bibitem{Nishiwaki2} Y. Nishiwaki {\it et al.}: submitted to J. Phys. Soc. Jpn.
\bibitem{Yoshizawa} H. Yoshizawa {\it et al.}: Phys. Rev. B {\bf 23} (1981) 2298.
\bibitem{Nagler} S. E. Nagler {\it et al.}: Phys. Rev. B {\bf 27} (1983) 1784.
\bibitem{Ishimura} N. Ishimura and H. Shiba: Progr. Theor. Phys. {\bf 63} (1980) 743.
\bibitem{Goff} J. P. Goff {\it et al.}: Phys. Rev. B {\bf 52} (1995) 15992.
\bibitem{Matsubara} F. Matsubara and S. Inawashiro: Phys. Rev. B {\bf 43} (1991) 796.
\bibitem{Shiba2} H. Shiba {\it et al.}: J. Phys. Soc. Jpn. {\bf 72} (2003) 2326.
\bibitem{Villain} J. Villain: Physica B {\bf 79} (1975) 1.
\bibitem{Shiba} H. Shiba: Progr. Theor. Phys. {\bf 64} (1980) 466.
\bibitem{Breitling} W. Breitling {\it et al.}: Solid State Comm. {\bf 24} (1977) 267.
\end{thebibliography}
\end{document}